\documentclass[floatfix,aps,twocolumn,prl,superscriptaddress]{revtex4-2}
\usepackage{graphicx}
\usepackage{dcolumn}
\usepackage{bm}
\usepackage[T1]{fontenc}
\usepackage{textcomp}
\usepackage{amsmath}
\usepackage{mathrsfs}
\usepackage{graphicx}
\usepackage{amssymb}
\usepackage{units}
\usepackage{xcolor}
\usepackage{float}

\begin{document}

\title{Lasing, quantum geometry and coherence   in 
non-Hermitian
flat bands}

\author{Ivan Amelio}
\affiliation{Center for Nonlinear Phenomena and Complex Systems,
Universit{\'e} Libre de Bruxelles, CP 231, Campus Plaine, B-1050 Brussels, Belgium}
\author{Nathan Goldman}
\affiliation{Center for Nonlinear Phenomena and Complex Systems,
Universit{\'e} Libre de Bruxelles, CP 231, Campus Plaine, B-1050 Brussels, Belgium}

\begin{abstract}
We show that lasing in flat band lattices can be stabilized by means of the geometrical properties of the Bloch states, in settings where the single-particle dispersion is flat in both its real and imaginary parts.  
We illustrate a general projection method and compute the collective excitations, which are shown to display a diffusive behavior ruled by quantum geometry through a peculiar coefficient involving gain, losses and interactions.  Then, we analytically show  that the phase dynamics display a surprising cancellation of the Kardar-Parisi-Zhang nonlinearity at the leading order.
Because of the relevance of Kardar-Parisi-Zhang universality in one-dimensional geometries, we focus our study on the diamond chain and provide confirmation of these results through full numerical simulations.
\end{abstract}

\date{\today}

\maketitle


\textit{Introduction.}
The physics of weakly dispersive bands, where correlation effects and interactions dominate over kinetic energy, has been the topic of various conceptual and experimental studies, starting from the strong coupling limit of the Hubbard model and the fractional quantum Hall effect, and pioneering, in recent years, the discovery of flat bands in twisted bilayer graphene~\cite{bistritzer2011} and the consequent experimental observation of superconductivity~\cite{cao2018super} and correlated insulators~\cite{cao2018correl}.
The most interesting flat bands are the ones 
displaying a complex structure of the Bloch states, not 
ascribable to a trivial atomic limit.
The crucial role of Bloch state geometry  for (quasi-)condensation~\cite{huber2010,takayoshi2013,julku2021} and superfluid transport~\cite{peotta2015,huhtinen2022revisiting}
has been pointed out.

In analogy to the case of weakly interacting atomic BECs~\cite{julku2021,Julku2023}, it is natural to consider lasing  in flat band systems in a semi-classical approximation.
In such a non-Hermitian context, however, one needs to specify whether the flatness condition applies to the real or imaginary part of the single-particle dispersion relation.
In this work, we are interested in studying the case where the Bloch geometry stabilizes a lasing state {\em in  a purely non-linear fashion}, as due to gain competition.
This goes beyond a number of recent results in topological lasers~\cite{noh2020,loirette-pelous2021} and 
polariton condensation
on the Lieb lattice~\cite{baboux2016}, where a privileged Bloch state enjoys a larger gain or quality factor {\em already at the linear level}.
In other words, we consider single-particle bands that are flat both in their real and imaginary parts.

Moreover, a question specific to non-equilibrium one-dimensional quasi-condenstates concerns the fate of Kardar-Parisi-Zhang (KPZ) physics~\cite{ji2015,he2015,squizzato2018,fontaine2022} in a flat band. Indeed, the KPZ nonlinearity is typically proportional to the bandwidth in the case of a dispersive band without quantum geometry.

This Letter starts by introducing the semi-classical lasing equations on the diamond chain and a real space projection method to the lowest flat band.
The geometry of the Bloch states  determines the steady-state lasing mode
and the collective modes, calculated by the Bogoliubov method.
We then go beyond Bogoliubov introducing stochastic noise and allowing for large phase fluctuations.
Adiabatically eliminating density fluctuations yields an equation for the phase with a full cancellation of a candidate nonlinear KPZ term, which is proportional to a Bloch geometric constant.
To check this remarkable conclusion, we perform numerical simulations
that display full agreement with the analytical argument.
Finally, we discuss connections with previous works on flat bands~\cite{baboux2016,Longhi2019,Harder2021}
and outline future directions.

\begin{figure}[t]
\centering
\includegraphics[width=0.35\textwidth]{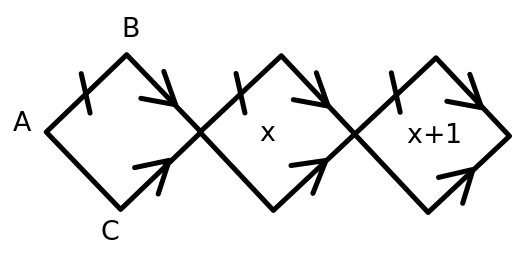}  
\caption{Sketch of the diamond chain, with three inequivalent sites $A,B,C$ per unit cell. The dash on the link stands for a minus sign in the hopping amplitude, the arrow for a $+i$.
}
\label{fig:diamond}
\end{figure}

\textit{Model and projection.}
We consider lasing in a semi-classical framework,
particularly adequate for
lattices of microring laser resonators~\cite{harari2018},
polariton micropillars~\cite{amo2016} and VCSEL's arrays~\cite{grabherr1999}. 
While a slow carrier dynamics can in practice give rise to further instabilities~\cite{longhi2018,baboux2018,Longhi2019,loirette-pelous2021}, here we assume that the adiabatic approximation holds.
The light field $\psi_{x\sigma}(t)$ on the lattice site $\sigma=1,...,N_\sigma$ in the $x$-th unit cell obeys then a complex Ginzburg Landau equation (CGLE)~\cite{carusotto2013}
\begin{multline}
    i\partial_t \psi_{x\sigma}
    =
    (H_0 \psi)_{x\sigma}
    +
    \\
    +
    \left\{
g |\psi_{x\sigma}|^2
+
\frac{i}{2} \left[P
\left(
1 - \frac{|\psi_{x\sigma}|^2}{n_S}
\right) - \gamma
\right]
    \right\}
    \psi_{x\sigma} ,
    \label{eq:CGLE}
\end{multline}
where $g$ is the refractive index nonlinearity, $P$ and $\gamma$ are respectively gain and losses, which we assume to be uniform along the system, $n_S$ is the saturation density that determines the strength of gain competition.
Finally,  the single particle Hamiltonian $H_0$
encodes the hopping on the lattice. 
Because of the special interest of 1D lattices for KPZ physics, we will be mainly dealing with the diamond (also known as rhombic) chain~\cite{vidal2000},
but we expect our analytical results to hold for other models and independently of the dimensionality of the system. 
The diamond chain is
sketched in Fig.~\ref{fig:diamond}
and is described by
\begin{equation}
    H_0 =
    -J \sum_{x=1}^{N_x}
     (c^\dagger_{x} a_{x}
     -b^\dagger_{x} a_{x}
    +i a^\dagger_{x+1} b_{x}
    +i a^\dagger_{x+1} c_{x})
    + \text{h.c.}
    ,
\end{equation}
where $\sigma$ runs over $A,B,C$ and we denoted 
$a \equiv \psi_{xA}, b \equiv \psi_{xB}, 
c \equiv \psi_{xC}$
 . The band flatness arises from destructive interference, since a flux $\pi$ pierces each plaquette; notice that for notational ease in what follows, we have chosen a gauge in which lasing will occur in the zero Bloch momentum state.
 Such Hamiltonian can be implemented with current technology in lattices of microring laser resonators~\cite{hafezi2011,bandres2018}. 

With periodic boundary conditions, the translationally invariant free particles are characterized by three perfectly flat bands with energies $\{ -2, 0, 2\}$.
In the following we assume that all the dynamics occurs in the lowest band, requiring the hopping $J$ to dominate over the other scales of the problem such as $P, \gamma$ and interactions. In practice one could have a gain medium spectrally centered on the lowest band but with a very large bandwidth.
The Bloch eigenstates corresponding to this band have the form 
$u_k(\sigma) = \frac{1}{2\sqrt{2}}
(2, -1 - ie^{ik}, 1 - ie^{ik})^T$, where $k$ denotes the quasi-momentum.
For a given field configuration,
its projection onto the lowest band has the form
$(\mathbb{P}\psi)_{x\sigma}
=
\sum_k u_k(\sigma) \frac{e^{ikx}}{\sqrt{N_x}} \bar{\psi}_k$.
Fourier transforming the coefficients $\bar{\psi}_k$
yields the auxiliary field
$\bar{\psi}_x$, which is the convenient representation to adopt in the following.
Indeed, by using the convolution
\begin{equation}
    (\mathbb{P}\psi)_{x\sigma}
=
\frac{1}{N_x} 
\sum_{k,y} u_k(\sigma) e^{ik(x-y)} \bar{\psi}_y ,
\end{equation}
the CGLE can be recast in the form 
\begin{multline}
    i\partial_t \bar{\psi}_x
    =
    i\frac{P-\gamma}{2}\bar{\psi}_x
    +
    \\
    +
\left( g - i\frac{P}{2n_S}
\right)
\sum_{y_1y_2y_3}
K(x,y_1,y_2,y_3)
\bar{\psi}^*_{y_1}
\bar{\psi}_{y_2}
\bar{\psi}_{y_3} ,
\label{eq:projectedCGLE}
\end{multline}
with the 
``quantum geometric
kernel''
\begin{equation}
    K(x,y_1,y_2,y_3)
    =
 \sum_{k_1k_2k_3}
    \frac{e^{ik_1z_1-ik_2z_2-ik_3z_3}}{N_x^3}
    \Lambda(k_1,k_2,k_3),
\end{equation}
where we used the shotcut $z_j\equiv y_j - x$ and introduced the crucial object
\begin{equation}
    \Lambda(k_1,k_2,k_3)
    =
    \sum_\sigma
    u^*_{k_2+k_3-k_1}(\sigma)
    u^*_{k_1}(\sigma)
    u_{k_2}(\sigma)
    u_{k_3}(\sigma),
\end{equation}
which contains the geometrical properties of the Bloch states.
For instance,
if $u_k = u^*_{-k}$,
and under the uniform density assumption 
$|u_0(\sigma)|^2=
\frac{1}{N_\sigma}$,
the quantity
$\sqrt{1-|\Lambda(k,0,0)|^2}$ represents the Hilbert-Schmidt distance between the Bloch states at $\pm k$~\cite{julku2021,Julku2023}.

In the following we will not be concerned with the original field and will drop the bar from $\bar{\psi}$.

\textit{Lasing state and Bogoliubov modes.}
So far we simply rewrote the CGLE projected onto the lowest band. Now we will study the steady-state solution and its dynamical properties.
Lasing will typically occur in the momentum state $k_*$ that optimizes gain saturation and interactions, i.e. the  momentum that minimizes $\Lambda(k,k,k)$.
In the case of the diamond chain it turns out that there exist two such momenta, 0 and $\pi$, and we will assume that $k_*=0$ is spontaneously selected. 
The lasing steady-state then reads
\begin{equation}
    \psi_x(t) = \sqrt{n_0} e^{-i\omega_0t} ,
\end{equation}
with $n_0=n_S \frac{P-\gamma}{P\Lambda_0 }$, $\omega_0 = gn_0\Lambda_0$
and the shortcut 
$\Lambda_0 \equiv \Lambda(0,0,0)$.

We now consider the collective modes on top of the steady-state by perturbing it as
$
\psi_x(t) = e^{-i\omega_0t} \left( \sqrt{n_0} +
\sum_k \delta\psi_k(t) e^{ikx}
\right)
$, obeying the equation of motion
\begin{multline}
    i\partial_t \delta\psi_k
    =
    (\mu - i\frac{\Gamma}{2})
    \left[
    (2\Lambda(0,k,0)-\Lambda_0) \delta\psi_k
    + \right.
    \\
    \left.
    + \Lambda(k,0,0) 
    \delta\psi^*_{-k}
    \right],
    \label{eq:Bogo_EOM}
\end{multline}
having defined $\mu\!=\!gn_0$
and $\Gamma\!=\!Pn_0/n_S$.
As noticed in previous studies on the collective excitations on top of an atomic BEC~\cite{julku2021,Julku2023}, the objects 
$\Lambda(0,k,0) = 
\sum_\sigma
    |u_{k}(\sigma)|^2
    |u_{0}(\sigma)|^2$
    and 
    $\Lambda(k,0,0) = 
\sum_\sigma
    u^*_{k}(\sigma) u^*_{-k}(\sigma)
    u^2_{0}(\sigma)$
are related to the quantum metric of the Bloch states
($\sqrt{1-|\Lambda(k,0,0)|^2}$ is also called the ``condensate quantum distance'').
We remark that, as a further improvement of these works, here we did not invoke any uniform density  assumption.
As usual, one can turn Eq.~(\ref{eq:Bogo_EOM}) into a $2\times2$ eigen-problem and find the complex eigenvalues $\omega_{\pm}(k)$. At $k=0$ the phase mode, called Goldstone mode, has always $\omega_+=0$.

\begin{figure}[t]
\centering
\includegraphics[width=0.4\textwidth]{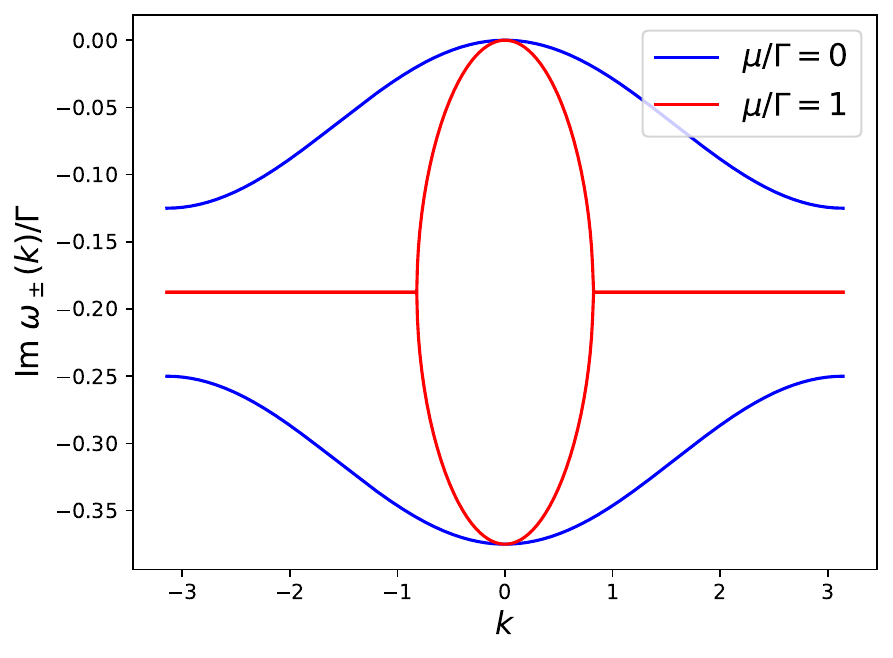}  
\caption{Imaginary part of the Bogoliubov modes for the diamond chain, see Eq.~(\ref{eq:Bogo_ws}), 
with and without interactions.
}
\label{fig:Bogo_diamond}
\end{figure}

In the case of the diamond chain,
$\Lambda(0,k,0)=\Lambda_0=\frac{3}{8}$
and, at small momentum,
$\Lambda(k,0,0)
\simeq \frac{3}{8} - \frac{\beta}{2} k^2$, with $\beta=1/8$ playing the role of a metric.
The Bogoliubov poles read
\begin{equation}
    \omega_{\pm}(k)
    =
    -\frac{i}{2}\Gamma \Lambda_0
    \pm
    \sqrt{
\mu^2 \Lambda_0^2
- (\mu^2 + \frac{\Gamma^2}{4})
\Lambda(k,0,0)^2
   + i0^+ },
   \label{eq:Bogo_ws}
\end{equation}
and for small momenta we recover for the Goldstone branch a diffusive behavior
\begin{equation}
    \omega_{+}(k) 
    \simeq
    -\frac{i}{2} \beta \frac{\mu^2 + \Gamma^2/4}{\Gamma/2} k^2
    \label{eq:Goldstone_branch}
    .
\end{equation}
We remind that for exciton-polaritons in dispersive bands~\cite{szymanska2006,wouters2007}, the Goldstone branch would diffuse as 
$\omega_{+}(k) 
    \simeq
    -i 4J \frac{\mu}{\Gamma} k^2$,
    so that not only Eq.~(\ref{eq:Goldstone_branch}) features the geometric quantity $\beta$, but also a peculiar functional dependence on $\Gamma, \mu$.
The Bogoliubov spectrum for the diamond chain is plotted in Fig.~\ref{fig:Bogo_diamond}.

\textit{Cancellation of the KPZ nonlinearity.}
So far we have been dealing with deterministic evolution and small perturbations.
We now supplement the CGLE 
(\ref{eq:CGLE})
with a stochastic drive $\sqrt{D} \xi_{x}(t)$ where 
$D$ is the strength of the noise 
and $\xi$ is taken as an uncorrelated random variable
of zero mean and unit variance, 
$\langle \xi_{x\sigma}^*(t)
\xi_{x'\sigma'}(t') \rangle = \delta_{xx'}\delta_{\sigma\sigma'} \delta(t-t')$.
Since $H_0$ is Hermitian and the Bloch states are orthogonal, when projecting onto the lowest band the noise variance is unaffected (in other words no Petermann broadening is introduced~\cite{amelio2022bogoliubov}), and one just needs to complement the projected CGLE (\ref{eq:projectedCGLE})
with a white noise term $\sqrt{D} \xi_{x}(t)$ of unit variance.

Because of the absence of long-range order, Bogoliubov theory in low dimensions is not fully consistent and its conclusions have to be taken with a grain of salt.
In one-dimensional non-equilibrium systems it has been established theoretically~\cite{ji2015,he2015,squizzato2018} and experimentally~\cite{fontaine2022} that the low-energy dynamics is dominated by phase fluctuations as described  by the KPZ equation~\cite{kardar1986}
\begin{equation}
    \partial_t \phi 
    =
    \nu \nabla^2 \phi + \lambda (\nabla\phi)^2
    + \sqrt{\mathcal{D}}
    \xi_{\phi},
\end{equation}
where 
$\langle \xi_{\phi}(xt) \xi_{\phi}(x't')\rangle
=
\frac{1}{2}\delta(x-x') \delta(t-t')$
and the KPZ nonlinearity
$\lambda$ differentiate this growth equation from the linear Gaussian evolution. While the exponential decay of the spatial correlations is unaffected by the presence of $\lambda$, the hallmark of KPZ physics is in the dynamical exponent characterizing the decay of temporal correlations, see below. 

To go beyond Bogoliubov theory, we adopt the density-phase formalism
$\psi(x,t)= \sqrt{n_0+\delta n} ~ e^{-i\omega_0 t + i \phi}$,
where we require only the density fluctuations to be small.
In the following we also imply that space has been coarse-grained and $x$ is now a continuous variable.

The major difference with the usual derivation of KPZ in polariton wires~\cite{gladilin2014} is that instead of a nonlocal but linear kinetic term and a local nonlinear term, in Eq.~(\ref{eq:projectedCGLE}) one has to deal with a {\em nonlocal and nonlinear} term.
To make progress we divide both sides of Eq.~(\ref{eq:projectedCGLE}) by $e^{i\phi}$ and expand
\begin{multline}
    e^{i(\phi(y_2)+\phi(y_3)-\phi(y_1)-\phi(x))} 
    \simeq 
    1 + i \nabla\phi(x) (z_2+z_3-z_1) +
    \\
    +  \frac{i}{2} \nabla\phi(x) (z^2_2+z^2_3-z_1^2)
    - \frac{1}{2}(\nabla\phi(x))^2 (z_2+z_3-z_1)^2.
\end{multline}
In principle, one can perform a Taylor expansion also for density fluctuations, but it turns out that the spatial derivatives of the density will eventually yield higher order corrections to the phase equation. For instance, the Laplacian $\nabla^2\delta n$ will generate a term like $\nabla^4 \phi$. Physically, this is related to the fact that density correlations are very short-ranged. We then just approximate  $\delta n(y_j) \simeq \delta n(x)$ (notice that  one drops $\nabla\delta n, \nabla^2\delta n$ also in the case of  dispersive bands).

\begin{figure}[t]
\centering
\includegraphics[width=0.45\textwidth]{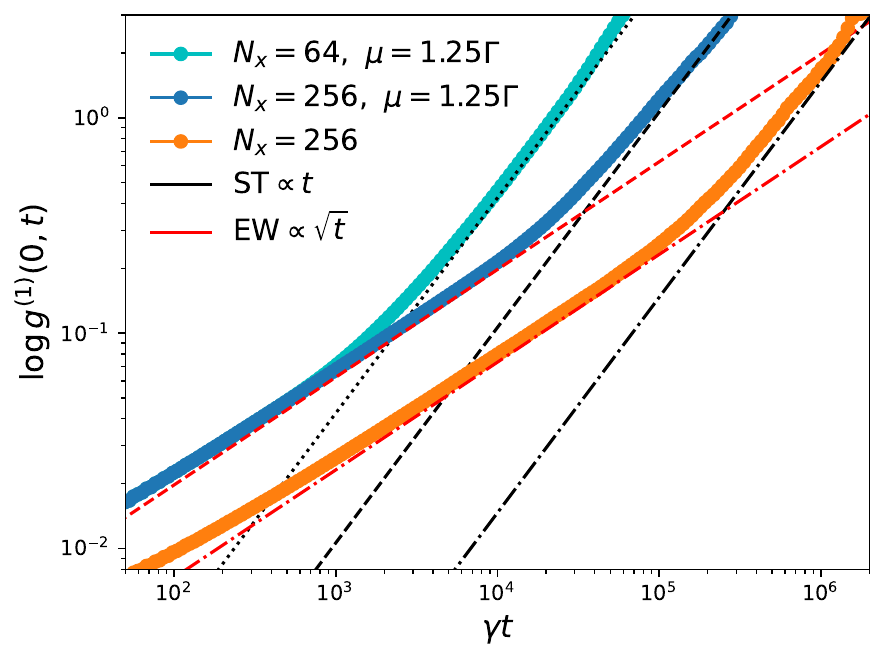} 
\includegraphics[width=0.45\textwidth]{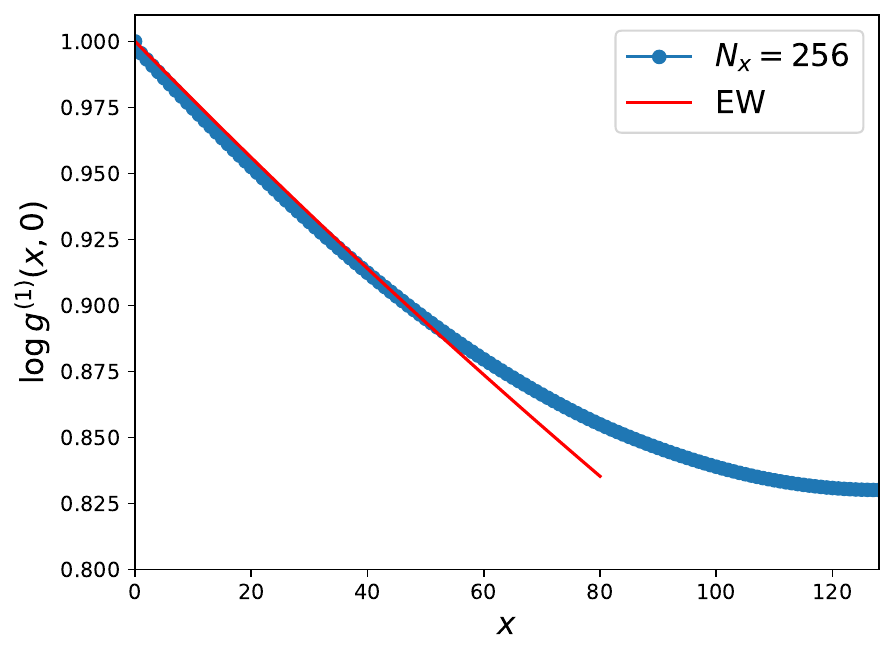}  
\caption{Correlation functions computed solving numerically Eq.~(\ref{eq:CGLE}) plus a stochastic drive.
(top) The temporal correlation function $g^{(1)}(0,t)$, displaying a crossover from Edwards-Wilkinson to Schawlow-Townes, and no sign of KPZ. The two EW red lines correspond to
$g=0$ (red dashed dotted) and $g=0.0005$ (red dashed), while the ST decay depends also on the size of the system (black dashed dotted  line for $g=0, N_x=256$, dashed and  for $g=0.0005, N_x=256$, dotted for $g=0.0005, N_x=64$). 
(bottom)
Exponential decay of $g^{(1)}(x,0)$, with the slope predicted by EW (before finite-size effects set in).
We take as parameters $P=2\gamma$, $n_S=500$, $D=\gamma$, $J=10\gamma$, and $g=0$ or $g=0.0005$, the latter corresponding to $\mu=1.25\Gamma$.
With this $J$, the overlap between the lasing mode and the linear Bloch state is $1-10^{-5}$,
ensuring that the projected theory works very well.
}
\label{fig:g1}
\end{figure}

We have reduced the problem to evaluating moments of the quantum geometric kernel
\begin{equation}
     \int dz_j \int dk_j ~
    e^{ik_1z_1-ik_2z_2-ik_3z_3}
    \Lambda(k_1,k_2,k_3) {\rm Pol}(z_1,z_2,z_3),
\end{equation}
where ${\rm Pol}(z_1,z_2,z_3)$ denotes a  polynomial with real coefficients.
In the case of the diamond chain, one can explicitly verify that 
$ \Lambda(k_1,k_2,k_3) =  \Lambda(-k_1,-k_2,-k_3)$.
This entails that only even polynomials survive the integration and the result
is real.
We denote $\eta$ the coefficient stemming in front of $(\nabla\phi)^2$,
while it is easy to recognize that integration of $z^2_2+z^2_3-z_1^2$
yields $\frac{d^2}{dk^2}\Lambda(k,0,0) = -\beta$.
Putting pieces together and separating real and imaginary parts of the CGLE, we finally arrive at the density-phase equations
\begin{equation}
    \partial_t\delta n
    =
    -\Gamma \Lambda_0 \delta n - \mu n_0 \beta \nabla^2 \phi + \frac{\Gamma}{2} n_0 \eta (\nabla \phi)^2 + 2 \sqrt{n_0 D} \xi_n
\end{equation}
\begin{equation}
    \partial_t \phi
    =
    -\mu \Lambda_0 \frac{\delta n}{n_0}
    + \frac{\Gamma}{4} \beta \nabla^2 \phi
    +\frac{\mu}{2} \eta (\nabla \phi)^2
    + \sqrt{\frac{D}{n_0}} \xi_\phi,
\end{equation}
where we now have two uncorrelated  sources of noise,
$\langle \xi_{\phi}(xt) \xi_{\phi}(x't')\rangle
=
\frac{1}{2}\delta(x-x') \delta(t-t')$
and
$\langle \xi_{n}(xt) \xi_{n}(x't')\rangle
=
\frac{1}{2}\delta(x-x') \delta(t-t')$.
The last step consists in adiabatically tracing out density fluctuations, so to be left with the equation for the low-energy phase dynamics
\begin{equation}
    \partial_t \phi =
    \frac{\beta}{2}  \frac{\mu^2 + \Gamma^2/4}{\Gamma/2} \nabla^2 \phi
    +
    \sqrt{\frac{D}{n_0}(1+(2\mu/\Gamma)^2)} ~ \xi_\phi.
    \label{eq:noKPZ}
\end{equation}
Remarkably, the KPZ nonlinearity is zero because of an exact cancellation of the $(\nabla \phi)^2$ term, that in principle had been generated by interactions and Bloch geometry.
Also, notice that the coefficient of the Laplacian recovers the diffusion of the Goldstone mode in Eq.~(\ref{eq:Goldstone_branch}) and the noise coefficient is the square root of the Henry-Schawlow-Townes linewidth per unit length~\cite{ameliochiocchetta}.

This argument for the cancellation of the KPZ nonlinearity holds irrespective of the dimensionality of the system, but it is of particular relevance in 1D, where the
renormalization group (RG) applied to the KPZ equation predicts that the Gaussian fixed point is unstable.
Then, in principle, the RG flow  may regenerate an effective KPZ nonlinearity originating from higher order corrections. Hence, we now aim at numerically testing the validity of our results by studying the presence of KPZ effects in the correlation functions.
We ran extensive simulations of the {\em full unprojected} CGLE (\ref{eq:CGLE})
plus stochastic drive, with an initial seed chosen to lase at $k\!=\!0$. We compute the statistical averages at the steady state
\begin{equation}
    g^{(1)}(x,t) = 
    \langle
    a_x^*(t) a_0(0) \rangle.
\end{equation} 
The hallmark of KPZ is a decay of temporal correlations of the form
$g^{(1)}(0,t) \sim \exp\left( - \mathcal{A} t^{2/3} \right)$ with some non-universal constant $\mathcal{A}$. In Fig. \ref{fig:g1}
we plot $-\log g^{(1)}(0,t)$ for different system sizes and values of $g$, showing no sign of the $2/3$ KPZ exponent.

On the other hand, Eq.~(\ref{eq:noKPZ}) has the form of a Gaussian process, described by the Edwards-Wilkinson (EW) equation
$\partial_t \phi 
    =
    \nu \nabla^2 \phi 
    + \sqrt{\mathcal{D}}
    \xi_{\phi}$.
    The correlations for this model can 
be easily computed and in 1D read
$
    g^{(1)}_{EW}(0,t) = n_0 \exp\left(-\frac{\mathcal{D}}{4} \sqrt{\frac{t}{\pi \nu}} \right)
     $ and  $
    g^{(1)}_{EW}(x,0) = n_0 \exp\left(-\frac{\mathcal{D}}{8\nu} x \right).
$
While these are the predictions for an infinite system, in general there will be finite size corrections. In particular, at large times we expect the Schawlow-Townes (ST) exponential decay
$g^{(1)}_{ST}(0,t)
= n_0 \exp\left(-\frac{\mathcal{D}}{4N_x } t \right)$, with linewidth inversely proportional to the size of the system~\cite{amelio2020theory}. 
We recall that the linewidth is also very sensitive
to the presence of KPZ physics, with a scaling $\sim N_x^{-1/2}$ that can show up already for a dozen of resonators~\cite{ameliochiocchetta}. 
The EW and ST predictions are also plotted in Fig. \ref{fig:g1}
and match perfectly the numerical data,
confirming the validity of Eq.~(\ref{eq:noKPZ}) and the absence of KPZ physics for all numerically accessible system sizes.
(The small discrepancy at small times is due to high-energy fluctuations for which density cannot be eliminated.)

\textit{Concluding remarks.}
We have revealed how gain competition  can stabilize lasing in flat bands with nontrivial Bloch geometry. This phenomenon occurs despite the fact that no mode is privileged at the linear level.

We have computed the Bogoliubov spectrum and the peculiar diffusion coefficient of the corresponding Goldstone branch.
Importantly, we have demonstrated that this flat-band setting leads to a total cancellation of the KPZ nonlinearity, such that flat-band lasers belong to a Gaussian universality class. This analytical result, which was obtained under weak assumptions, was numerically validated using the diamond chain. An intriguing question concerns the fate of this Gaussian behavior and the Schawlow-Townes prediction for the linewidth in the limit of arbitrarily large systems.

Previous works based on the Lieb chain~\cite{baboux2018} considered a non-uniform gain to induce lasing in a flat band, which led to a momentum-dependent gain $P(k) = P_0 - P_2 k^2 + ...$ at the linear level determined by the shape of the Bloch states. In that case, a KPZ nonlinear term proportional to $P_2$ is known to appear in the equation for the phase~\cite{he2015}. Instead, in this work we have  considered bands that are flat in both their real and imaginary part, which led to a cancellation of the KPZ term.
It would be interesting to consider a situation where one mode is favoured at the linear level, while another mode optimizes gain saturation.

We notice that a PT-symmetric flatband laser was proposed in \cite{Longhi2019} and a Kagome polariton condensate was realized in \cite{Harder2021},
without analyzing the role of quantum geometry.
In both these cases, the flat band is not gapped from the dispersive bands, such that our projected theory does not apply. 
It would be interesting to assess the presence of the KPZ nonlinearity in these cases.

Here we limited ourselves to a semi-classical theory, where interactions are weak enough to approximate the field on each site by a coherent state.
It would be of great interest to investigate the quantum regime of strong interactions and explore flat band physics in
quantum dissipative systems.
Experimental platforms like circuit QED promise to be very well suited for this purpose~\cite{martinez2023flatband,kolovsky2023}. We remark that clean samples are needed, since flat bands are particularly sensitive to disorder~\cite{baboux2016}.

\textit{Acknowledgements.}
We are grateful to Daniele De Bernardis, Maxime Burgher, Sebastian Diehl and Stefano Longhi for stimulating  discussions and exchanges.
This research was financially supported by the ERC grant LATIS, the EOS project CHEQS and the FRS-FNRS (Belgium). All numerical calculations were performed using the Julia Programming Language \cite{bezanson2017}.


\bibliography{bibliography}

\end{document}